\documentstyle[12pt]{article}

\renewcommand{\baselinestretch}{1.65}

\newtheorem{theorem}{Theorem}            

\begin{document}

\renewcommand{\baselinestretch}{1.1}

\title{\Large \bf Riemann-Cartan Space-times 
          of G\"odel Type  \\} 

\author{
J.E. {\AA}man\thanks{ {\sc ja@physto.se} },  \ \  
J.B. Fonseca-Neto\thanks{
{\sc  jfonseca@dfjp.ufpb.br}  }, \ \ 
M.A.H. MacCallum\thanks{ {\sc M.A.H.MacCallum@qmw.ac.uk} }, \ \ \\
\&  \  
M.J. Rebou\c{c}as\thanks{
{\sc  reboucas@cat.cbpf.br}  }  \\ 
\\  
$^{\ast}~$Institute of Theoretical Physics, 
             Stockholm University   \\ 
             Box 6730 (Vanadisv\"agen 9),
             S-113\ 85 \ \ Stockholm, Sweden \\
\\
$~^{\dagger}~$Departamento de F\'{\i}sica,
             Universidade Federal da Para\'{\i}ba  \\
             Caixa Postal 5008, 58059-900 Jo\~ao Pessoa -- PB, Brazil \\   
\\
$^{\ddagger}~$School of Mathematical Sciences, 
              Queen Mary \& Westfield College \\
              Mile End Road,
              London E1 4NS, U.K. \\ 
\\
$^{\S}$~Centro Brasileiro de Pesquisas F\'\i sicas\\
        Departamento de Relatividade e Part\'\i culas \\
        Rua Dr.\ Xavier Sigaud 150,
       22290-180 Rio de Janeiro -- RJ, Brazil \\
        }
        
\date{}  

\maketitle

\begin{abstract}
A class of Riemann-Cartan G\"odel-type space-times are examined in 
the light of the equivalence problem techniques. The conditions
for local space-time homogeneity are derived, generalizing previous
works on Riemannian G\"odel-type space-times. 
The equivalence of Riemann-Cartan G\"odel-type space-times of this 
class is studied. It is shown that they admit a five-dimensional group
of af\/fine-isometries and are characterized by three essential 
parameters $\,\ell, m^2, \omega$: identical triads ($\ell, m^2, \omega$) 
correspond to locally equivalent manifolds.
The algebraic types of the irreducible parts of the curvature and 
torsion tensors are also presented. 
\end{abstract}

{\raggedright
 \section{Introduction} }      \label{intro}
\setcounter{equation}{0}

Theories with non-zero torsion have been of notable interest in
a few contexts. In the framework of gauge theories they have
been used in the search for unification of gravity with the other
fundamental interactions in physics. Space-time manifolds with 
nonsymmetric connection have also been considered as the appropriate 
arena for the formulation of a quantum gravity theory. At a classical 
level the generalization of the standard general relativity by 
introducing torsion into the theory has also received a good deal
of attention mainly since the sixties (see Hehl {\em et al.\/}%
~\cite{Hehl1976} and references therein). The geometric concept of 
torsion has also been used in a continuum approach to lattice defects 
in solids since the early fifties%
~\cite{Kroener1981}~--~\cite{Balachandran1997}.
For further motivation and physical consequences of 
studying manifolds and theories with non-zero torsion (whether 
dropping the metricity condition or not) as well as for 
a detailed list of references on theories with non-zero torsion, 
we refer the readers to a recent review by Hehl 
{\em et al.\/}~\cite{Hehl1995}.

In general relativity (GR), the space-time  $M$ is a 
four-dimensional Riemannian manifold endowed with a 
locally Lorentzian metric and a metric-compatible 
symmetric connection, namely the Christof\/fel symbols 
$\{_{b\ c}^{\ a}\}$.
However, it is well known that the metric tensor and the 
connection can be introduced as independent structures on a 
given space-time manifold $M$. 

In GR there is a unique torsion-free connection on $M$. 
In the framework of torsion theories of gravitation (TTG),
on the other hand, we have Riemann-Cartan (RC) manifolds, 
i.e., space-time manifolds endowed with locally 
Lorentzian metrics and metric-compatible nonsymmetric 
connections ${\Gamma}^a_{\ bc}$. Thus, in TTG the 
connection has a metric-independent part given by 
the torsion, and for a characterization of the local 
gravitational field, one has to deal with both metric 
and connection. 

The arbitrariness in the choice of coordinates is a basic 
assumption in GR and in TTG. 
Nevertheless, in these theories it gives rise to the 
problem of deciding whether or not two apparently dif\/ferent 
space-time solutions of the field equations are locally the 
same --- the equivalence problem. In GR this problem can be
couched in terms of local isometry, whereas in TTG 
besides local isometry ($g_{ab} \rightarrow \tilde{g}_{ab}$) 
it means af\/fine collineation~%
($\Gamma^{a}_{\ bc} \rightarrow \tilde{\Gamma}^{a}_{\ bc}$) 
of two RC manifolds.

The equivalence problem in general relativity (Riemannian 
space-times) has been discussed by several authors and is of 
interest in many contexts (see, for example, Cartan~\cite{cartan},
Karlhede~\cite{karl}, MacCallum~\cite{mm1}~--~\cite{MacCSkea}
and references therein).
 
The equivalence problem in torsion theories of gravitation
(RC space-times), on the other hand, was only discussed 
recently~\cite{frt}. Subsequently,  an algorithm for checking 
the equivalence in TTG  and  a first working version 
of a computer algebra package (called {\sc tclassi}) which
implements this algorithm have been presented~\cite{frm}~--~\cite{frm1}.

The G\"{o}del~\cite{godel} solution of Einstein's field equations 
is a particular case of the G\"{o}del-type line element, defined
by 
\begin{equation}
ds^{2} = [ dt + H(x)\, dy]^{2} - D^{2}(x) \, dy^{2} - dx^{2} - dz^{2},
                                                        \label{ds2}
\end{equation}
in which 
\begin{equation}
  H(x) = e^{m x}, \;\;\;\;  D(x) = e^{m x}/ \sqrt{2},    \label{gddgod}
\end{equation}
and with the en\-ergy\--mo\-men\-tum tensor $T_{\mu \nu}$ given by 
\begin{eqnarray}
& T_{\mu \nu}=\rho v_{\mu} v_{\nu}\,, \qquad 
  v^{\alpha}=\delta^{\alpha}_{\ 0}\,,&  \label{gdsrc1} \\
& \kappa \rho = - 2 \Lambda = m^{2} = 2\, \omega^{2}\,, \label{gsol} & 
\end{eqnarray}  
where $\kappa$ and $\Lambda$ are, respectively, the Einstein gravitational
and the
cosmological constants, $\rho$ is the fluid density and $v^{\alpha}$ its
four-velocity, and $\omega$ is the rotation 
of the matter. The G\"{o}del model is homogeneous in space-time 
(hereafter called ST homogeneous).
Actually it admits a five parameter group of isometries ($G_{5}$) having
an isotropy subgroup of dimension one ($H_{1}$).

The problem of space-time homogeneity of  four-dimensional 
Riemannian manifolds endowed with a G\"{o}del-type metric (\ref{ds2})
was considered for the first time by Raychaudhuri and 
Thakurta~\cite{raytha}. They have determined the necessary conditions 
for space-time homogeneity. 
Afterwards, Re\-bou\-\c{c}as and Tiom\-no~\cite{rebtio} proved 
that the Ray\-chau\-dhuri-Tha\-kurta necessary conditions are also 
suf\/ficient for ST homogeneity of G\"{o}del-type Riemannian space-time 
manifolds. However, in both articles~\cite{raytha,rebtio} the
study of ST homogeneity is limited in that only time-independent
Killing vector fields were considered~\cite{tra}. 
The necessary and suf\/ficient conditions for a 
G\"{o}del-type Riemannian space-time manifold to be ST homogeneous 
were finally rederived without assuming such a simplifying hypothesis 
in~\cite{rebaman}, where the equivalence problem techniques for 
Riemannian space-times, as formulated by Karlhede~\cite{karl} and 
implemented in {\sc classi}~\cite{Aman}, were used.

In this article, in the light of the equivalence problem techniques 
for Riemann-Cartan space-times, as formulated by Fonseca-Neto 
{\em et al.}~\cite{frt} and embodied in the suite of computer algebra 
programs {\sc tclassi}~\cite{frm}~--~\cite{frm1}, we shall examine 
all Riemann-Cartan manifolds endowed with a G\"odel-type metric
(\ref{ds2}), with a torsion polarized along the preferred direction 
defined by the rotation, and sharing the translational symmetries of 
$g_{\mu\nu}$ in (\ref{ds2}). Hereafter, for the sake of brevity,
we shall refer to this family 
of space-time manifolds as Riemann-Cartan G\"odel-type manifolds.
The necessary and suf\/ficient conditions for a Riemann-Cartan 
G\"odel-type  manifold to be ST (locally) homogeneous are derived.
The {\AA}man-Rebou\c{c}as results~\cite{rebaman} for Riemannian
G\"odel-type space-times are generalized.
The ST homogeneous Riemann-Cartan G\"odel-type manifolds are shown to 
admit a five-dimensional group of af\/fine-isometric motions. The equivalence
of these Riemann-Cartan space-times is discussed: they are found to be 
characterized by three essential parameters $m^2$, $\omega$ and $\ell$: 
identical triads ($\ell, m^2, \omega$) correspond to equivalent manifolds.
The algebraic classification of the nonvanishing irreducible 
parts of the curvature is presented. For a general triad 
($\ell, m^2, \omega$), the Weyl-type spinors $\Psi_A$ and 
$\psi_A$ are both Petrov type D, while the non-null Ricci-type 
spinors  $\Phi_{AB'}$ and $\phi_{AB'}$ are both Segre type 
[1,1(11)]. 
A few main instances for which these algebraic types can be
more specialized are also studied. The classification of
the irreducible parts of the torsion and the corresponding
group of isotropy are also discussed.
The pseudo-trace torsion spinor ${\cal P}_{AX'}$ is found 
to be space-like, with $SO(2,1)$ as its group of isotropy. 
The Lanczos spinor ${\cal L}_{ABCX'}$ is found to be invariant 
under one-dimensional spatial rotation.

Our major aim in the next section is to present a brief summary
of some important theoretical and practical results on the 
equivalence problem for Riemann-Cartan space-times required in 
Section 3, where we state, prove and discuss our main results.

\vspace{5mm}

{\raggedright
\section{Equivalence of Riemann-Cartan Space-times: Basic Results} }
\setcounter{equation}{0}  \label{Equivalence}

\begin{sloppypar}
Most relativists's first approach to solving the (local)
equivalence problem of Riemann-Cartan manifolds would probably
be to make use of the so-called scalar polynomial invariants
built from the curvature, the torsion, and their covariant
derivatives~\cite{Christensen1980}. However, this attempt
cannot work since there exist {\em curved} plane
wave RC space-times with non-zero torsion~\cite{Adamowicz1980} 
for which all the scalar polynomial
invariants vanish --- indistinguishable therefore from
the Minkowski space (flat and torsion free). This example
shows that although necessary the scalar polynomial invariants
are not suf\/ficient to distinguish (locally) two RC
space-times.  \end{sloppypar}

To make apparent that the conditions for the local equivalence
of RC manifolds follow from Cartan's approach to the 
equivalence problem, we shall first recall the definition of
equivalence and then proceed by pointing out how Cartan's
results~\cite{cartan} lead to the solution of
the problem found in~\cite{frt}.
The basic idea is that if two Riemann-Cartan manifolds $M$ and 
$\widetilde{M}$ are the same, they will define identical
Lorentz frame bundles [$L(M) \equiv L(\widetilde{M})$].
The manifold $L(M)$ incorporates the freedom in the choice of 
Lorentz frames and has a uniquely-defined set of linearly
independent 1-forms $\{ \Theta^{A}, \omega^{A}_{\ B} \}$, forming a
basis of the cotangent space $T^{\ast}_{P}(L(M))$ at an arbitrary
point $P \in L(M)$. 
Two RC manifolds $M$ and $\widetilde {M}$ are then said to be
locally equivalent when there exists a local mapping $F$ 
between the Lorentz frame bundles $L(M)$ and $L(\widetilde{M})$ 
such that (see~\cite{frt} and also Ehlers~\cite{Ehlers1981})  
\begin{equation} \label{equidef}
F^{\ast}\,\widetilde{\Theta}^{A} = \Theta^{A}   
\qquad \mbox{and} \qquad
F^{\ast}\,\widetilde{\omega}^{A}_{\ B} = \omega^{A}_{\ B}   
\end{equation}
hold. Here $F^{\ast}$ is the well known pull-back map
defined by $F$.

A solution to the equivalence problem for Riemann-Cartan 
manifolds can then be obtained by using Cartan's results
on the equivalence of sets of 1-forms (see p.\ 312 of the English translation 
of Ref.~\cite{cartan}) together with 
Cartan equations of structure for a manifold endowed with a
nonsymmetric connection. The solution can 
be summarized as follows~\cite{frt,frm1}. 

Two $n$-dimensional Riemann-Cartan (locally Lorentzian) manifolds $M$ and 
$\widetilde{M}$ are locally equivalent if there exists a local map 
(diffeomorphism) $F$ between their corresponding Lorentz frame 
bundles $L(M)$ and $L(\widetilde{M})$, such that the 
{\em algebraic} equations relating the components 
of the curvature and torsion tensors and their covariant derivatives:
\begin{eqnarray}  
\label{eqvcond} 
T^{A}_{\ BC}    & = & \widetilde{T}^{A}_{\ BC}\;, \nonumber \\
R^{A}_{\ BCD} & = &  \widetilde{R}^{A}_{\ BCD}\;, \nonumber \\
T^{A}_{\ BC;M_{1}}  & = &  \widetilde{T}^{A}_{\ BC;M_{1}}\;, \nonumber \\ 
R^{A}_{\ BCD;M_{1}} & = & \widetilde{R}^{A}_{\ BCD;M_{1}}\;, \nonumber  \\
T^{A}_{\ BC;M_{1}M_{2}}  & = &  \widetilde{T}^{A}_{\ BC;M_{1}M_{2}}\;,  \\ 
                  & \vdots &   \nonumber \\
R^{A}_{\ BCD;M_{1}\ldots M_{p+1}} & = & \widetilde{R}^{A}_{\ BCD;M_{1}
                                             \ldots M_{p+1}}\;,\nonumber \\ 
T^{A}_{\ BC ;M_{1} \ldots M_{p+2}} & = & \widetilde{T}^{A}_{\ BC;M_{1} 
                                   \ldots M_{p+2}} \nonumber 
\end{eqnarray} 
are compatible as equations in Lorentz frame bundle coordinates
$\left( x^{a}, \xi^{A} \right)$. Here 
and in what follows we use a semicolon to denote covariant derivatives.
Note that $x^{a}$ are coordinates on the manifold $M$ while $ \xi^{A}$
parametrize the group of allowed frame transformations. Reciprocally, 
equations (\ref{eqvcond}) imply local equivalence between the space-time 
manifolds. The $(p+2)^{th}$ derivative of torsion and the $(p+1)^{th}$
derivative of curvature are the lowest derivatives which are functionally 
dependent on all the previous derivatives. 
It should be noticed that in the above set of {\em algebraic\/} equations
{\em necessary and sufficient\/} for the local equivalence we have taken 
into account the Bianchi identities $R^{A}_{\ \ [\,BCD\,]}
-T^{A}_{\ \ [\,BC;D\,]}  = - T^{N}_{\ \ [\,BC}T^{A}_{\ D\,]\,N}$
and their dif\/ferential concomitants. Thus, when the components of 
the $\,0^{th}, \ldots ,(p+1)^{th}\,$ covariant derivatives of 
the torsion are known, the Bianchi identities and their dif\/ferential
concomitants reduce to a set of linear algebraic equations, which
relates (for each $p$) the $(p+1)^{th}$ covariant derivatives of curvature 
to the $(p+2)^{th}$ covariant derivatives of torsion. 
So we need the $(p+2)^{th}$ derivatives of torsion in (\ref{eqvcond}), 
which did not appear in~\cite{frt}.

A comprehensive local description of a Riemann-Cartan manifold 
is, therefore, given by the set 
\begin{equation}
I_{p} = \{ T^{A}_{\ BC}\,, R^{A}_{\ BCD}\,, T^{A}_{\ BC;N_{1}}\,, 
 R^{A}_{\ BCD;M_{1}}\,,T^{A}_{\ BC;N_{1}N_{2}}\,, \,\ldots, \, 
 R^{A}_{\ BCD;M_{1} \ldots M_{p}\,,} T^{A}_{\ BC;N_{1} \ldots N_{p+1}} \},  
\label{rcscl}
\end{equation}
whose elements are called Cartan scalars, since they 
are scalars under coordinate transformations on the base manifold.
The theoretical upper bound for the number of covariant 
derivatives to be calculated is $10$ for the curvature and
$11$ for the torsion, which corresponds to $11$ steps (from
$0$th to $10$th-order derivatives for the curvature) 
in the algorithm presented below. 
The number of steps can be thought of as being related to 
the six Lorentz transformation parameters $\xi^A$, 
the four coordinates $x^a$ on the space-time manifold 
and one integrability condition.
A word of clarification regarding this integrability condition 
is in order here: when the number of derivatives needed is not 
the maximum possible (set by the dimension of the frame bundle) 
then to show that the derivative process has terminated one has to 
take one more derivative and show that it contains
no new information by checking the functional relations
between the Cartan scalars. This can be understood as if 
we were introducing invariantly-defined coordinates
(though we cannot explicitly do that) and then had to take
their derivatives in order to substitute for the dif\/ferentials 
in the usual formula for the line element. 
In practice, the coordinates and Lorentz transformation
parameters are treated dif\/ferently. 
Actually a fixed frame (a local section of the Lorentz frame bundle)
is chosen to perform the calculations 
so that the elements of the set $I_{p}$ coincide with the the components 
of the curvature and torsion tensors of the space-time base manifold 
and their covariant derivatives; there is no explicit 
dependence on the Lorentz parameters.	

To deal with  equivalence it is necessary to
calculate the elements of the set $I_{p}$. However,
even when the Bianchi and Ricci identities and their 
dif\/ferential concomitants are taken into account, in the 
worst case one still has 11064 independent elements to 
calculate. Thus, an algorithmic procedure for carrying out 
these calculations and a computer algebra implementation
are highly desirable. 

A practical procedure for testing equivalence of Riemann-Cartan 
space-times has been  developed~\cite{frm}~--~\cite{frm1},
\cite{afmr1}. In the procedure 
the maximum order of derivatives is not more than $7$ for the curvature 
and $8$ for the torsion.
The basic idea behind our procedure is separate handling of frame 
rotations and space-time coordinates, fixing the frame at each 
stage of dif\/ferentiation of the curvature and torsion tensors by 
aligning the basis vectors as far as possible with 
invariantly-defined directions.
The algorithm starts by setting $q=0$ and has the following 
steps~\cite{frm1}:
\begin{itemize}
\begin{enumerate}
\item 
Calculate the set $I_{q}$, i.e.,  
the derivatives of the curvature up to the $q^{th}$ order 
and of the torsion up to the $(q+1)^{th}$ order.
\item 
Fix the frame, as much as possible, by putting
the elements of $I_{q}$ into canonical forms.
\item 
Find the frame freedom given by the residual
isotropy group $H_{q}$ of transformations which
leave the canonical forms invariant.
\item 
Find the number $t_{q}$ of functionally 
independent functions of space-time coordinates 
in the elements of $I_q$, brought into the canonical 
forms.
\item   
If the isotropy group $H_{q}$ is the same
as  $H_{(q-1)}$ and the number of functionally independent
functions $t_{q}$ is equal to $t_{(q-1)}$,
then let $q=p+1$ and stop. Otherwise, increment 
$q$ by 1 and go to step $1$.    
\end{enumerate}
\end{itemize}

This procedure provides a discrete characterization of 
Riemann-Cartan space-times in terms of the following 
properties: the set of canonical forms in $I_{p}$, 
the isotropy groups $\{H_{0},\ldots ,H_{p}\}$ and the 
number of independent functions $\{t_{0}, \dots ,t_{p}\}$. 
Since there are $t_p$  essential space-time coordinates,
clearly $4-t_p$ are ignorable, so the isotropy group
will have dimension  $s = \mbox{dim}\,( H_p )$, and the group of 
symmetries (called af\/fine-isometry) of both metric (isometry) 
and torsion (af\/fine collineation) will have dimension $r$ given by
\begin{equation}
r = s + 4 - t_p \,, \label{gdim}
\end{equation}
acting on an orbit with dimension
\begin{equation}
d = r - s = 4 - t_p \,.  \label{ddim}
\end{equation}

To check the equivalence of two Riemann-Cartan space-times
one first compares the above  discrete properties and only when
they match is it necessary to determine the compatibility
of equations (\ref{eqvcond}).

In our implementation of the above practical procedure, rather
than using the curvature and torsion tensors as such, the 
algorithms and computer algebra programs were devised and 
written in terms of spinor equivalents:
(i) the irreducible parts of the Riemann-Cartan curvature spinor
\begin{eqnarray}
 R_{ABCDG'H'} & = & \varepsilon_{G'H'} \, [ \Psi_{ABCD}  
 +(\varepsilon_{AC}\varepsilon_{BD} +
   \varepsilon_{AD}\varepsilon_{BC})(\Lambda + i\Omega)  
 +  \varepsilon_{AC}\Sigma_{BD} \nonumber \\ 
&+& \varepsilon_{BD}\Sigma_{AC} 
 + \varepsilon_{AD}\Sigma_{BC} 
 + \varepsilon_{BC}\Sigma_{AD} \,]
 + \varepsilon_{CD}(\Phi_{ABG'H'} + i\Theta_{ABG'H'})\,,  \label{spcurv}
\end{eqnarray}
which, clearly, are
$\Psi_{ABCD}$ ({\sc tpsi}),  $\Phi_{ABX'Z'}$ ({\sc tphi}),
$\Theta_{ABX'Z'}$  ({\sc theta}), $\Sigma_{AB}$ ({\sc sigma}), 
$\Lambda$ ({\sc tlambd}) and $\Omega$ ({\sc omega}); and (ii) the 
irreducible parts of the torsion spinor
\begin{equation}
 T_{AX'BC} = L_{X'ABC} +\frac{1}{3}\,(\varepsilon_{AB}T_{CX'} +
   \varepsilon_{AC}\bar{T}_{BX'}) 
 + \frac{1}{3}\,i\,(\varepsilon_{AB}S_{CX'} +
   \varepsilon_{AC}\bar{S}_{BX'})\,,  \label{sptor}
\end{equation}
namely: ${\cal T}_{AX'}$ ({\sc spttor}, {\sc sp} $=$ spinor,
{\sc t} $=$ trace, {\sc tor} $=$ torsion ), 
${\cal P}_{AX'}$ ({\sc spptor}, {\sc sp} $=$ spinor, {\sc p} $=$
pseudo-trace, {\sc tor} $=$ torsion ), 
and ${\cal L}_{ABCX'}$ ({\sc spltor}, {\sc sp} $=$ spinor,
{\sc l} $=$ Lanczos spinor, {\sc tor} $=$ torsion). 
Note that these irreducible parts of curvature and torsion spinors 
are nothing but the spinor equivalents of the curvature and torsion
tensors given, respectively, by equations (B.4.3) and (B.2.5) 
in the appendix B of Ref.~\cite{Hehl1995}.
Note that the {\sc tclassi} users'  names for the spinorial quantities 
have been indicated inside round brackets. 
We note that, besides the above indication for the names of the
irreducible parts of the torsion spinor,  the names of the 
irreducible parts of both Riemann-Cartan curvature and first covariant 
derivative of the torsion were generalized 
from the names in {\sc classi}~\cite{MacCSkea} by bearing in mind whether 
they have the same symmetry as the Weyl spinor (the Weyl-type 
spinors: $\Psi_A$ and $\psi_A$) or the symmetry of the Ricci spinor 
(the Ricci-type spinors: $\Phi_{AB'}$, $\phi_{AB'}$, 
$\Theta_{AB'}$, $\nabla {\cal T}_{AX'}$, $\nabla {\cal P}_{AY'}$).
We have employed the af\/fixes: {\sc bv} for bivector, {\sc sp} for 
spinor, {\sc sc} for scalar, {\sc a} for d'Alembertian;
{\sc d}, {\sc d2} and so on, for the first, 
the second and so forth derivative of the spinorial quantities. 
We have used the letters $\Sigma$, ${\cal M}$, ${\cal B}$ 
to denote bivectors, i.e., objects with the same symmetries of Maxwell
spinor. We have also named three basic scalars by {\sc tlambd} ($\Lambda$)
{\sc omega} ($\Omega$) and {\sc scttor} (${\cal T}$). There are also
names which were simply borrowed from {\sc classi} with the addition
of the letter {\sc t} for torsion, as for example {\sc txi}
(see~\cite{MacCSkea}, \cite{frm} and \cite{frm1} for details).

A relevant point to be taken into account when one
needs to compute derivatives of the curvature and the
torsion tensors is that they are interrelated by
the Bianchi and Ricci identities and their dif\/ferential concomitants.
Thus, to cut down the number of quantities to be calculated
it is very important to find a minimal set of quantities from
which the curvature and torsion tensors, and their covariant
derivatives are obtainable by algebraic operations.
For Riemann-Cartan space-time manifolds, 
taking into account the irreducible parts of the Bianchi and 
Ricci identities and their differential concomitants, 
a complete minimal set $C_q$ of
such quantities recursively defined in terms of totally symmetrized
$q^{th}$ derivatives of the curvature spinors
and $(q+1)^{th}$ derivatives of the torsion 
spinors can be specified~\cite{frm1,fmr2} by: 
\begin{enumerate} 
\item
 For $q=0$ : the torsion's irreducible parts, namely  \\
(a) ${\cal T}_{AX'}$ ({\sc spttor}), (b) ${\cal P}_{AX'}$ ({\sc spptor}), 
(c) ${\cal L}_{ABCX'}$  
({\sc spltor});
\item    The totally symmetrized $q^{th}$ derivatives of
   \begin{enumerate}
   \item 
      \begin{enumerate} 
      \item $\Psi_{ABCD}$ ({\sc tpsi}),
      \item $\psi_{ABCD} \equiv - \nabla^{N'}_{\ \ \ (A}{\cal L}^{}_{BCD)N'}$ 
           ({\sc psiltor}),
      \end{enumerate} 
   \item 
      \begin{enumerate} 
      \item $\Phi_{ABX'Z'}$ ({\sc tphi}),
      \item $\Theta_{ABX'Z'}$  ({\sc theta}),
      \item $\phi_{ABX'Z'} \equiv - \frac{1}{2} (
       \nabla^{N}_{\ \ (X'}{\cal L}^{}_{Z')ABN} + 
       \nabla^{N'}_{\ \ (A}{\bar {\cal L}}^{}_{B)X'Z'N'} ) $ ({\sc philtor}), 
      \end{enumerate} 
   \item 
      \begin{enumerate}  
      \item  $\Lambda$ ({\sc tlambd}),
      \item  $\Omega$  ({\sc omega}),
      \item ${\cal T} \equiv \nabla_{NN'}{\cal T}^{NN'}$ ({\sc scttor}),
      \end{enumerate} 
   \item 
      \begin{enumerate}
      \item $\Sigma_{AB}$  ({\sc sigma}),
      \item ${\cal M}_{AB} \equiv \nabla^{N'}_{\ \ \ (A}{\cal T}^{}_{B)N'}$ 
          ({\sc bvttor}),   
      \item ${\cal B}_{AB} \equiv \nabla^{N'}_{\ \ \ (A}{\cal P}^{}_{B)N'}$ 
          ({\sc bvptor})\,.  
      \end{enumerate}
\vspace{2mm}
    \end{enumerate} 
\item     The totally symmetrized $(q+1)^{th}$ derivatives of
  (a) ${\cal T}_{AX'}$ ({\sc dspttor}), 
  (b) ${\cal P}_{AX'}$ ({\sc dspptor}), 
  (c) ${\cal L}_{ABCX'}$ ({\sc dspltor}).
\item     For $q \geq 1$:
   \begin{enumerate}
   \item 
   the totally symmetrized $(q-1)^{th}$  derivatives of
      \begin{enumerate}
      \item 
$\Xi_{ABCX'} \equiv \nabla^{N}_{\ X'}\Psi^{}_{ABCN}\,$ 
      ({\sc txi}),
      \item 
${\cal X}_{ABCX'} \equiv \nabla^{N'}_{\ (A}\Theta^{}_{BC)N'X'}\,$ 
      ({\sc xith}),  
      \item 
${\cal U}_{AX'} \equiv \frac{1}{2}(\nabla_{\ \ X'}^{N} \Sigma_{AN} 
            +\nabla_{\ \ A}^{N'}{\bar \Sigma}_{X'N'})\,$  ({\sc tsigm}),
      \item 
${\cal V}_{AX'} \equiv -\frac{i}{2}(\nabla_{\ \ X'}^{N} \Sigma_{AN}
              -\nabla_{\ \ A}^{N'}{\bar \Sigma}_{X'N'})$  ({\sc psigm});
      \end{enumerate}
   \item            
for $q=1$ the d'Alembertian of the irreducible parts of the torsion: 
       \begin{enumerate}
       \item 
$\Box\,{\cal T}_{AX'} \equiv \nabla^{NN'}\nabla_{NN'}\,{\cal T}_{AX'}$ 
                     ({\sc aspttor}),
       \item 
$\Box\,{\cal P}_{AX'} \equiv \nabla^{NN'}\nabla_{NN'}\,{\cal P}_{AX'}$ 
                     ({\sc aspptor}), 
       \item 
$\Box\,{\cal L}_{ABCX'} \equiv 
                \nabla^{NN'}\nabla_{NN'}\,{\cal L}_{ABCX'}$ ({\sc aspltor}).
       \end{enumerate}
   \end{enumerate}
\item       For $q \geq 2$:
    \begin{enumerate}
    \item
the d'Alembertian $\Box\,Q \equiv \nabla^{NN'}\nabla_{NN'}\,Q$ applied 
to all quantities $Q$ calculated for the derivatives of order $q-2$,
i.e. in the set $C_{(q-2)}$, except the
d'Alembertians of the irreducible parts of torsion for $q=2$ 
(when $n=2$, e.g., $\,\Box\, \Psi_A$ ({\sc atpsi}), 
$\Box\, \psi_A$ ({\sc apsiltor}),
$\Box\, \Phi_{AB'}\,$ ({\sc atphi}), and so forth).
    \item
the totally symmetrized $(q-2)^{th}$ derivatives of
        \begin{enumerate}
        \item 
$ \Upsilon_{ABCD} \equiv -\nabla^{N'}_{\ \ (A}{\cal X}^{}_{BCD)N'}\,$  
     ({\sc psixith}),  
       \item 
${\cal F}_{AB} \equiv \nabla^{N'}_{\ \ \ (A}{\cal U}^{}_{B)N'}\,$   
     ({\sc bvtsigm}). 
      \end{enumerate}
   \end{enumerate} 
\end{enumerate}

It should be stressed that we have included in the above set the
d'Alembertian of the irreducible parts of the torsion (4.(b)~i~--~iii), 
which was missed in~\cite{frm}.
Note also that the above list contains inside parentheses the
{\sc tclassi} external name (for the users) after each quantity.
Finally, we remark that the above minimal set is a generalization 
of the corresponding set found for Riemannian space-time manifolds 
by MacCallum and {\AA}man~\cite{MacAman}.

In our practical procedure the frame is fixed (as much as possible)
by bringing into canonical form first the quantities with the 
same symmetry as the Weyl spinor (called Weyl-type), i.e., $\Psi_A$ 
and $\psi_A$, followed by the spinors with the symmetry of the 
Ricci spinor (referred to as Ricci-type spinor), namely
$\Phi_{AB'}$,  $\phi_{AB'}$, $\Theta_{AB'}$, $\nabla {\cal T}_{AX'}$, 
$\nabla {\cal P}_{AY'}$, then
bivector spinors $\Sigma_{AB}$, ${\cal M}_{AB}$ and ${\cal B}_{AB}$, 
and finally vectors ${\cal T}_{AX'}$ and ${\cal P}_{AX'}$
are taken into account. Thus, if $\Psi_{A}$ is Petrov I, for example, 
the frame can be fixed by demanding that the nonvanishing components 
of $\Psi_A$ are such that $\Psi_1 = \Psi_3 \not= 0, \Psi_2 \not= 0$. 
Clearly an alternative canonical frame is obtained by imposing 
$\Psi_0 = \Psi_4 \not= 0, \Psi_2 \not= 0$. Although the latter is
implemented in {\sc tclassi} as the canonical frame, in the next section 
we shall use the former (defined to be
an acceptable alternative in {\sc tclassi}) 
to make easier the comparison between our results and those of the 
corresponding Riemannian case~\cite{rebaman}. 

To close this section we remark that in the {\sc tclassi} 
implementation of the above results a notation is used  in 
which the indices are all subscripts and components are labelled
by a primed and unprimed index whose numerical values are the
sum of corresponding (primed and unprimed) spinor indices. 
Thus, for example, one has
\begin{equation}
\nabla\,\Psi_{20'} \equiv  \Psi_{(1000;1)0'} =
\nabla^{X'}_{\ \ \ (A} \Psi^{}_{BCDE)}\,\iota^A \iota^B o^C o^D o^E\, 
            \bar{o}_{X'}\,, \nonumber
\end{equation}
where the parentheses indicate symmetrization, the bar is
used for complex conjugation, and the pair
($\iota^A, o^B$) constitutes an orthonormal spinor basis.

\vspace{5mm}

{\raggedright
\section{Homogeneous Riemann-Cartan G\"odel-type  Space-times} }
\setcounter{equation}{0} 

Throughout this section we shall consider a four-dimensional
Riemann-Cartan manifold $M$, endowed with a G\"odel-type 
metric (\ref{ds2}) and a torsion that shares the same translational
symmetries as the metric, and is aligned with the direction 
singled out by the rotation vector field $w$ (called
Riemann-Cartan G\"odel-type space-time).
So, in the coordinate system given in (\ref{ds2}) the torsion 
tensor is given by $T^t_{\ xy} = S(x)$.

For arbitrary functions $H(x)$, $D(x)$ and $S(x)$ both $\Psi_A$
and $\psi_A$ are Petrov I; this fact can be easily checked by 
using the package {\sc tclassi}. Accordingly the null tetrad $\Theta^A$
which turns out to be appropriate (canonical) for our discussions
is
\begin{eqnarray}
\Theta^{0} = \frac{1}{\sqrt{2}}(\theta^{0} + \theta^{3})\,, \quad\qquad
\Theta^{1} = \frac{1}{\sqrt{2}}(\theta^{0} - \theta^{3})\,, \nonumber \\
\label{nullt} \\
\Theta^{2} = \frac{1}{\sqrt{2}}(\theta^{2} - i \theta^{1})\,, \quad\qquad
\Theta^{3} = \frac{1}{\sqrt{2}}(\theta^{2} + i \theta^{1})\,,   \nonumber
\end{eqnarray}
where $\theta^{A}$ is a Lorentz tetrad
($\eta_{AB} = {\rm diag}\,(+1,-1,-1,-1)$) given by
\begin{equation}
\theta^{0} = dt + H(x)\,dy\,, \quad
\theta^{1} = dx\,, \quad
\theta^{2} = D(x)\,dy\,, \quad
\theta^{3} = dz\,.          \label{lort}
\end{equation}
Clearly in the null frame (\ref{nullt}) the torsion tensor and the
G\"odel-type line element (\ref{ds2}) are given by
\begin{equation}
T^0_{\ 23} = T^{1}_{\ 23} = \frac{\sqrt{2}}{2}\,i\,S(x)  \qquad 
\mbox{and} \qquad                      
ds^2 = 2\,(\Theta^0\,\Theta^1 - \Theta^2\,\Theta^3)\,.  \label{gtyrc}
\end{equation}
It is worth mentioning that the Petrov type for $\Psi_A$ and $\psi_A$
and the canonical frame (\ref{nullt}) were obtained by interaction with
{\sc tclassi}, starting from the Lorentz frame (\ref{lort}), changing
to a null tetrad frame, and making dyad transformations to bring
$\Psi_A$ and $\psi_A$ into the canonical form for Petrov type I
discussed in section 2.

Using the {\sc tclassi} package we referred to in the previous sections
one finds the following nonvanishing components of the Cartan scalars 
corresponding to the first step (for $q=0$) of our algorithm:
\begin{eqnarray}
\Psi_1 &=& \Psi_3 = \frac{1}{8}\,\left[\,S' - \left( \frac{H'}{D}\, \right)' 
                         \, \right] \,,     \label{1st}  \\
\Psi_2 &=& - \,\frac{S}{4}\, \left(\frac{S}{3} - \frac{H'}{D}\, \right) 
+\frac{1}{6}\left[\,\frac{D''}{D} - \left( \frac{H'}{D}\, \right)^2\, 
           \,\right] \,, \\  
\psi_1 &=& \psi_3 = \frac{S'}{8} \,, \\ 
\psi_2 &=& - \,\frac{S}{4}\,\left( S - \frac{H'}{D}\, \right) \,,  \\ 
\Phi_{00'}&=&\Phi_{22'} = \frac{S}{4}\, \left(\frac{S}{2} 
-\,\frac{H'}{D}\, \right) +\frac{1}{8}\, \left(\frac{H'}{D}\,\right)^2 \,,\\ 
\Phi_{01'}&=&\Phi_{12'} = - \,\frac{S'}{8} 
           + \frac{1}{8}\, \left(\frac{H'}{D}\,\right)' \,, \\
\Phi_{11'}&=& \,\frac{S}{4}\,\left(\,\frac{S}{4} - \frac{H'}{D}\, \right)  
  + \frac{1}{4} \left[\, \frac{3}{4}\,\left(\frac{H'}{D}\,\right)^2 
  - \,\frac{D''}{D} \right] \,, \\
\phi_{00'}&=& \phi_{22'} = \phi_{11'} = \frac{S}{4}\, 
                     \left( S - \frac{H'}{D}\,\right) \,, \\ 
\phi_{01'}&=& \phi_{12'} = - \,\frac{S'}{8} \,, \\
\nabla {\cal P}_{01'} &=& -\,\nabla {\cal P}_{12'} =
         -\,\frac{i}{4}\, S' \,, \\
{\cal B}_0 &=& - {\cal B}_2 = - \,\frac{i}{2}\, S' \,, \\ 
{\cal P}_{00'}&=& - {\cal P}_{11'} = - \,\frac{\sqrt{2}}{2}\, S \,, \\ 
\Lambda &=& - \frac{S^2}{48}\,- \frac{1}{12} \left[\,\frac{D''}{D}
   - \frac{1}{4} \left(\frac{H'}{D}\,\right)^2  \right] \,,  \\
{\cal L}_{10'}&=& {\cal L}_{21'} = - \,\frac{i}{6}\,\sqrt{2}\,S \,, \\
\nabla {\cal L}_{10'}&=&- \,\nabla {\cal L}_{32'}= \frac{S}{16}\,
                \left( S -\frac{H'}{D}\, \right)\,, \\
\nabla {\cal L}_{11'}&=&- \,\nabla {\cal L}_{31'}= 
   -\,\frac{3}{4} \,\nabla {\cal L}_{20'}=  
    \frac{3}{4} \,\nabla {\cal L}_{22'} = \frac{S'}{16} \,,  \label{last}
\end{eqnarray}
where the prime denotes derivative with respect to $x$.

{}From equation (\ref{ddim})  one  finds that for ST 
homogeneity we must have $t_{p} = 0$, that is the number 
of functionally independent functions of the space-time 
coordinates in the set $I_{p}$ must be zero. Accordingly all 
the above  quantities of the  minimal set must be constant. 
Thus, from  eqs.\ (\ref{1st}) -- (\ref{last}) one easily concludes 
that for a Riemann-Cartan G\"odel-type space-time (\ref{gtyrc}) to 
be ST homogeneous it is necessary that
\begin{eqnarray}
S  &=& \mbox{const} \equiv \ell \,,  \label{torcond} \\
\frac{H'}{D} &=& \mbox{const} \equiv 2\,\omega \label{metcond1} \,, \\
\frac{D''}{D}&=& \mbox{const} \equiv m^2 \,. \label{metcond2}
\end{eqnarray}

We shall now show that the above necessary conditions are also
suf\/ficient for ST homogeneity. Indeed, under the conditions
(\ref{torcond}) -- (\ref{metcond2}) the Cartan scalars 
corresponding to the first step (for $q=0$) of our algorithm reduce to
\begin{eqnarray} 
\Psi_2 &=&  \,\frac{\ell}{2}\, \left(\,\omega - \frac{\ell}{6} \,\right)
    +\frac{m^2}{6} - \frac{2}{3}\,\omega^2 \,, \label{um} \\
\psi_2 &=&-\,\frac{\ell}{4}\,\left(\ell -2\,\omega\,\right)\,,\label{dois} \\ 
\Phi_{00'}&=&\Phi_{22'} =\frac{\ell}{4}\,\left(\frac{\ell}{2} 
 -2\,\omega\,\right) +\frac{\omega^2}{2} \,, \label{tres} \\
\Phi_{11'}&=& \,\frac{\ell}{4}\,\left(\,\frac{\ell}{4} - 2\,\omega\, \right)  
  +  \frac{3}{4}\,\omega^2 - \frac{m^2}{4} \,, \label{quatro} \\
\phi_{00'}&=& \phi_{22'} = \phi_{11'} = \frac{\ell}{4}\, 
                \left( \ell - 2\,\omega\,\right) \,, \label{cinco} \\
{\cal P}_{00'}&= & -{\cal P}_{11'} =
            -\,\frac{\sqrt{2}}{2}\, \ell\,,\label{seis} \\ 
\Lambda &=& -\frac{\ell^2}{48}\,+\frac{1}{12} \left(\,\omega^2-m^2 \right)\,,
                                        \label{sete} \\
{\cal L}_{10'}&=& {\cal L}_{21'} =
              -\,\frac{i}{6}\,\sqrt{2}\,\ell\,,\label{oito}\\
\nabla {\cal L}_{10'}&=&- \,\nabla {\cal L}_{32'}= \frac{\ell}{16}\,
                \left( \ell - 2\,\omega\,\right)\,. \label{nove}
\end{eqnarray}

Following the algorithm of the previous section, one needs to find
the isotropy group which leaves the above Cartan scalars (canonical
forms) invariant. Since $\ell \not= 0$ one can easily find that 
there are Cartan scalars invariant under the three-dimensional Lorentz
group $SO(2,1)$ like, e.g., ${\cal P}_{AB'}$, or even the whole Lorentz 
group, like $\Lambda$. However, the whole set of Cartan 
scalars (\ref{um})~--~(\ref{nove}) is invariant only under the spatial 
rotation
\begin{equation} \label{SpaRot}
 \left( 
\begin{array}{cc}
   e^{i\alpha} &      0       \\
        0      & e^{- i\alpha} \\
\end{array} 
\right) \,\,,
\end{equation}
where $\alpha$ is a real parameter. So, the residual group $H_0$ which 
leaves the above Cartan scalars invariant is one-dimensional.

We proceed by carrying out the next step of our practical procedure,
i.e., by calculating the totally symmetrized covariant derivative of 
the Cartan scalars (\ref{um})~--~(\ref{nove}) and the d'Alembertian 
of the irreducible parts of the torsion.
Using {\sc tclassi} one finds the following nonvanishing 
quantities:
\begin{eqnarray} 
\nabla \,\Psi_{20'} &=& -\nabla \,\Psi_{31'} =
\frac{i}{40}\,\sqrt{2}\,\,\ell \left(2\,m^2 + 8\,\ell\, \omega
   - \ell^2 - 20\,\omega^2 \,\right) \nonumber \\
&& 
+\frac{i}{10}\,\sqrt{2}\,\omega \left( 4\,\omega^2 - m^2\right)\,,
                                            \label{dez} \\
\nabla \,\psi_{20'} &=& -\nabla \,\psi_{31'} =
 -\frac{i}{40}\,3\,\sqrt{2}\,\,\ell \left( 4\,\omega^2
   - 4\,\ell\, \omega + \ell^2 \,\right)\,,    \label{onze} \\
\Xi_{10'} & = & \Xi_{21'} =
 \frac{i}{16}\,\sqrt{2}\,\,\ell \left( 2\,m^2 + 8\,\ell\, \omega 
   - \ell^2 - 20\, \omega^2\, \right)
\nonumber \\
&&
+\frac{i}{4}\,\sqrt{2}\,\omega\left( 4\,\omega^2 - m^2 \right) 
                                             \label{doze} \,, \\
\Box\,{\cal L}_{10'}&=& \,\Box\,{\cal L}_{21'}= 
\frac{i}{4}\,\sqrt{2}\,\,\ell \left(4\,\omega^2 
   - 4\,\ell\, \omega + \ell^2 \,\right)\,,    \label{treze}  \\
\nabla^2 {\cal L}_{10'}&=& \,\nabla^2 {\cal L}_{43'}= 
 \frac{i}{80}\,\sqrt{2}\,\,\ell \left( 4\,\omega^2 
   - 4\,\ell\, \omega + \ell^2 \,\right)\,,  \label{quatorze}  \\
\nabla^2 {\cal L}_{21'}&=& \,\nabla^2 {\cal L}_{32'}= 
-\,\frac{i}{480}\,\sqrt{2}\,\,\ell \left(4\,\omega^2 
   - 4\,\ell\, \omega +\ell^2 \,\right)\,.     \label{quinze} 
\end{eqnarray} 

As no new functionally independent function arose, $t_0=t_1$. 
Besides, the Cartan scalars (\ref{dez}) -- (\ref{quinze}) are 
invariant under the same isotropy group (\ref{SpaRot}), 
i.e. $H_0 = H_1$. Thus no new covariant derivatives 
should be calculated. From eq.\ (\ref{gdim}) one finds that 
the group of symmetries (af\/fine-isometric motions) of the 
Riemann-Cartan G\"odel-type space-time is five-dimensional --- the 
necessary conditions (\ref{torcond})~--~(\ref{metcond2}) 
are also suf\/ficient for ST homogeneity. 

The above results can be collected together in the following theorems:

\vspace{2mm} 
\begin{theorem} \label{HomCond}
The necessary and sufficient conditions for a Riemann-Cartan
G\"odel-type space-time to be ST (locally) homogeneous 
are those given by  equations (\ref{torcond})~--~(\ref{metcond2}).
\end{theorem}
\begin{theorem} \label{RelPar}   \begin{sloppypar}
All ST locally homogeneous Riemann-Cartan G\"o\-del-type space-\-times
admit a five-dimensional group of affine-isometric motion and
are characterized by three independent parameters $\,\ell$,
$m^2$ and $\omega$: identical triads ($\ell, m^2, \omega$) specify 
equivalent space-times.              \end{sloppypar}
\end{theorem}

As the parameter $\omega$ is known to be essentially the rotation
in G\"odel-type space-times, a question which naturally arises here 
is whether there is any simple geometrical interpretation for the 
parameters $\ell$ and $m$. As the parameter $\ell$ is a measure
of the strength of the torsion it clearly has the geometrical 
interpretation usually associated with the torsion tensor.
We have not been able to figure out a simple geometrical interpretation
for the parameter $m$, though. The parameter $m^2$, nevertheless, has 
been used to group the general class of G\"odel-type metrics into three
disjoint subclasses, namely: (i) the hyperbolic class ($m^2 > 0$),
(ii) the circular class ($m^2 \equiv -\,\mu^2 < 0$), and the
linear class, where $m^2=0$ (see in this regard~\cite{rebtio}).

It is worth emphasizing that when $\ell = 0$  eqs.\ (\ref{um})~--~%
(\ref{quinze}) reduce to the corresponding equations for Riemannian 
G\"odel-type space-times (eqs.\ (3.12)~--~(3.15) and (3.18)~--~(3.21) 
in~\cite{rebaman}). Therefore, the results in~\cite{rebaman} can be 
reobtained as a special case of our study here. So, for example, the above 
theorems \ref{HomCond} and \ref{RelPar} generalize the corresponding 
theorems in~\cite{rebaman} (theorems 1 and 2 on page 891).

It should be noticed that the Riemannian ST-homogeneous G\"odel-type 
space-times can have a group of isometries of dimension higher
than five, as, e.g.,  when $m^2 = 4\,\omega^2$ which permits
a $G_7$, and $\omega = 0$, $m \not= 0$ which allows a $G_6$. 
However, for Riemann-Cartan G\"odel-type space-times, apart from 
the rather special case of flat Riemann-Cartan space-time 
($\ell=m=\omega=0$), there are no relations among 
the relevant parameters ($\ell, m^2, \omega$) for which the 
dimension of the group of af\/fine-isometric motions is higher 
than five. 

As far as the algebraic classification of the nonvanishing 
Weyl-type and Ricci-type spinors  is concerned, from 
eqs.\ (\ref{um}) -- (\ref{cinco}) we find that for a general
triad $(\ell, m^2, \omega)$ both Weyl-type spinors $\Psi_A$ and 
$\psi_A$ are Petrov type D, whereas the Ricci-type spinors 
$\Phi_{AB'}$ and $\phi_{AB'}$ are both of Segre type [1,1(11)]. 
There exist, nevertheless, many instances for which these algebraic 
types can be more specialized. We mention a few:
\begin{enumerate}
\item 
When either $m= \ell/3 = 2\,\omega$ or $m^2=\ell^2/2,\;\;\omega=0$,
$\Psi_A$ and $\psi_A$ are, respectively, Petrov 0 and D, while
both $\Phi_{AB'}$ are Segre type [(1,11)1]; and both $\phi_{AB'}$ 
are type [1,1(11)]; 
\item 
For $\ell = 2\,\omega$ and $m \not= 0$, $\Psi_A$ is Petrov D, 
$\psi_A$ is Petrov 0, $\Phi_{AB'}$ is type [(1,1)(11)], and 
$\phi_{AB'}$ is Segre type 0; 
\item
When $\ell = 2\,\omega$ and $m=0$, both $\Psi_A$ and $\psi_A$ 
are Petrov 0, and  $\Phi_{AB'}$ and $\phi_{AB'}$ are both Segre 
type 0;
\item
For $m=\,\omega= 0,\;\; \ell \not=0$ (Riemannian flat space-time),
$\Psi_A$ and  $\psi_A$ are both Petrov type D, while
$\Phi_{AB'}$ is Segre type [1,(111)]; and $\phi_{AB'}$ 
is type [1,1(11)].
\end{enumerate}

Regarding the classification of the nonvanishing parts of 
the torsion spinors one can easily find that for $\ell \not=0$ 
the spinor ${\cal P}_{AX'}$ corresponds to a space-like vector, 
with $SO(2,1)$ as its group of isotropy. The Lanczos spinor 
${\cal L}_{ABCX'}$ is invariant under the spatial rotation~%
(\ref{SpaRot}) (one-dimensional isotropy group). 

It should be noticed that the equivalence problem techniques,
as formulated in Ref.~\cite{frt} and embodied in the suite of
computer algebra programs~{\sc tclassi} which we have used in 
this work, can certainly be used in more general contexts, as
for example in the examination of Riemann-Cartan G\"odel-type 
family space-times in which the torsion, although polarized 
along the direction of the rotation, does not share the 
translational symmetries of the metric~\cite{fr}. 
We have chosen the case of the present article
because it gives a simple illustration of our approach to
the equivalence problem techniques applied to Einstein-Cartan
G\"odel-type solutions which have already been discussed in the 
literature (see Ref.~\cite{DTT2} and references 
therein quoted on G\"odel-type solutions with torsion).

As well as specialist systems such as {\sc sheep}, on which
{\sc tclassi} is based, all the main general-purpose computer algebra
systems have some sort of facilities for calculation in general
relativity. Indeed, extensive sets of programs useful in GR are
available with {\sc reduce}, {\sc maple} and {\sc macsyma}, and with
{\sc mathematica} through the {\sc MathTensor} package. By contrast,
as far as we are aware, the existing facilities in computer algebra
systems for calculations in theories with non-zero torsion are quite
limited.
Actually, we only know of the {\sc reduce} programs for applications 
to Poincar\'e gauge field theory written by J. Dermott McCrea
in collaboration with F. W. Hehl~\cite{McCrea} and a set of 
{\sc mathematica} programs for calculation in RC manifolds written 
by H. H. Soleng and called {\sc cartan}~\cite{Soleng96} 
(see also~\cite{Hehl97}).
McCrea's programs are written using the {\sc reduce} package 
{\sc excalc}~\cite{McCrea,Schrufer}. These programs, however, 
do not contain the implementation of the equivalence problem for
Riemann-Cartan manifolds. The {\sc lisp}-based system {\sc tclassi} was 
devised with the equivalence problem of RC manifold in mind, and is so 
far the only package that incorporates the equivalence problem techniques 
(see also~\cite{frm} and \cite{frm1}).
Furthermore, also in TTG there is room for specialized systems like
{\sc tclassi}. The major reason for this is that they tend to be 
more ef\/ficient than general-purpose systems. 
For a comparison of CPU times for a specific metric in GR, for
example, see MacCallum~\cite{mm2,mm3}.

To conclude, we should like to emphasize that as no field
equations were used to show the above results, they are valid 
for every Riemann-Cartan G\"odel-type solution
regardless of the torsion theory of gravitation one is
concerned with, in particular they hold for the Riemann-Cartan
G\"odel-type class of solutions discussed in~\cite{DTT2} 
and~\cite{DTT1}, which were found in the context of Einstein-Cartan
theory.

\vspace{4mm}
\section*{Acknowledgments}
J.B. Fonseca-Neto and M.J. Rebou\c{c}as gratefully acknowledge 
financial assistance from CNPq.

\end{document}